\documentclass{PoS}
\usepackage{graphicx}
\usepackage[percent]{overpic}
\usepackage{caption}
\usepackage{gensymb}
\usepackage{subcaption}
\usepackage{url}
\newcommand{\overlaploc}{\mathrm{Overlap}_\mathrm{loc}}
\newcommand{\overlapext}{\mathrm{Overlap}_\mathrm{ext}}

\newcommand{\apj}{ApJ}
\newcommand{\apjs}{ApJS}
\newcommand{\hi}{$\mathrm{H\,\scriptstyle{I}}$}
\newcommand{\g}{$\gamma$}
\newcommand{\st}{$^{\mathrm{st}}$}  
\newcommand{\nassocprobclassified}{$36$}
\newcommand{\nmock}{two} 
\newcommand{\nmockN}{2}  
\newcommand{\nmockpercent}{$\sim6\%$} 
\newcommand{\nninetyfivemock}{eight} 
\newcommand{\nninetyfivemockpercent}{$22\%$} 
\newcommand{\nclassifiedsnrs}{$30$} 

\newcommand{\nmarginalAIEM}{two} 
\newcommand{\nGalSNRs}{$279$} 
\newcommand{\nnotsnrs}{four} 
\newcommand{\ndetected}{$102$} 
\newcommand{\nextended}{$17$} 
\newcommand{\nnewpointlike}{$10$} 
\newcommand{\nnewextended}{four} 
\newcommand{\Fermi}{\emph{Fermi}}  
\newcommand{\FermiLat}{\emph{Fermi} LAT}     

\title{Systematically characterizing regions of the First \FermiLat~ SNR Catalog}

\ShortTitle{Systematically characterizing regions of the First Fermi-LAT SNR Catalog}

\author{
\speaker{F.~de Palma}$^{\;\; \dagger \,(1,2)\;}$, 
T.~J.~Brandt$^{\;(3)\;}$, 
J.~W.~Hewitt$^{\;(3,4)\;}$, 
G.~Johannesson$^{\;(5)\;}$,
L.~Tibaldo$^{\;(6)\;}$
on behalf of the \FermiLat~ Collaboration \\
$^{(1)}$ Istituto Nazionale di Fisica Nucleare, Sezione di Bari, 
Bari, Italy \\
$^{(2)}$ Universit\'a Telematica Pegaso, Piazza Trieste e Trento, 48, 80132 Napoli, Italy \\
$^{(3)}$ NASA Goddard Space Flight Center, Greenbelt, MD 20771, USA \\
$^{(4)}$ CRESST/University of Maryland, Baltimore County, Baltimore, MD 21250, USA \\
$^{(5)}$ Science Institute, University of Iceland, Dunhaga 3, 107 Reykjavik, Iceland \\
$^{(6)}$ Kavli Institute for Particle Astrophysics \& Cosmology, Slac National Accelerator Laboratory, USA \\
$\dagger$ {\footnotesize{E-mail:}} {\tt{\footnotesize{francesco.depalma@ba.infn.it}}}}

\abstract{While supernova remnants (SNRs) are widely thought to be powerful cosmic-ray accelerators, indirect evidence comes from a small number of well-studied cases. Here we systematically determine the gamma-ray emission detected by the \Fermi~ Large Area Telescope (LAT) from all known Galactic SNRs, disentangling them from the sea of cosmic-ray generated photons in the Galactic plane. Using LAT data we have characterized the 1-100 GeV emission in 279 regions containing SNRs, accounting for systematic uncertainties caused by source confusion and instrumental response. We have also developed a method to explore some systematic effects on SNR properties caused by the modeling of the interstellar emission (IEM). The IEM contributes substantially to gamma-ray emission in the regions where SNRs are located. To explore the systematics we consider different model construction methods, different model input parameters, and independently fit the model components to the gamma-ray data. We will describe this analysis method in detail. In the First \FermiLat~SNR Catalog there are 30 sources classified as SNRs, using spatial overlap with the radio position. For all the remaining regions we evaluated upper limits on SNRs' emission. In this work we will present a study of the aggregate characteristics of SNRs, such as comparisons between GeV and radio sizes as well as fluxes and spectral indexes and with TeV.}

\FullConference{The 34th International Cosmic Ray Conference,\\
		30 July- 6 August, 2015\\
		The Hague, The Netherlands}

\begin{document}
\section{Introduction}
In this work we will shortly describe the procedure used in the systematic search and analysis of supernova remnants (SNR) in the \Fermi-Large Area Telescope (LAT) data. The complete procedure will be described in a future publication \cite{snrcat}. Several results of this work will be described and highlighted also in \cite{ICRC_talk}. 
\section{Analysis procedure}
Our analysis starts by considering the 274 radio SNRs described in \cite{2009BASI...37...45G} and five new SNRs discovered later. Using their radio position and size we analyze three years of Pass7v6 ``Source''-class \FermiLat~ data in the energy range between 1 and 100 GeV. For each of these radio SNRs we test their possible gamma-ray counterpart extension, location, spectral curvature and significance. The method used for the spectral and spatial fit is described in \cite{Lande12-extSrcSearch}. Since we don't know exactly the background sources a priori for all these sources we developed a method that iteratively adds background sources on top of the standard interstellar emission model (IEM) gal\_2yearp7v6\_v0.fits, and isotropic model iso\_p7v6source.txt. As background sources we used also the pulsar described in the second pulsar catalog \cite{2PC} and the associated sources in the LAT second source catalog \cite{2FGL}. For all this candidate SNRs we evaluated their spectral characteristics and location, if we did not detect any source we reported a 99\% Bayesian upper limit (UL) \cite{Helene83} .

\begin{figure}
\centering
\begin{overpic}[width=0.52\columnwidth]{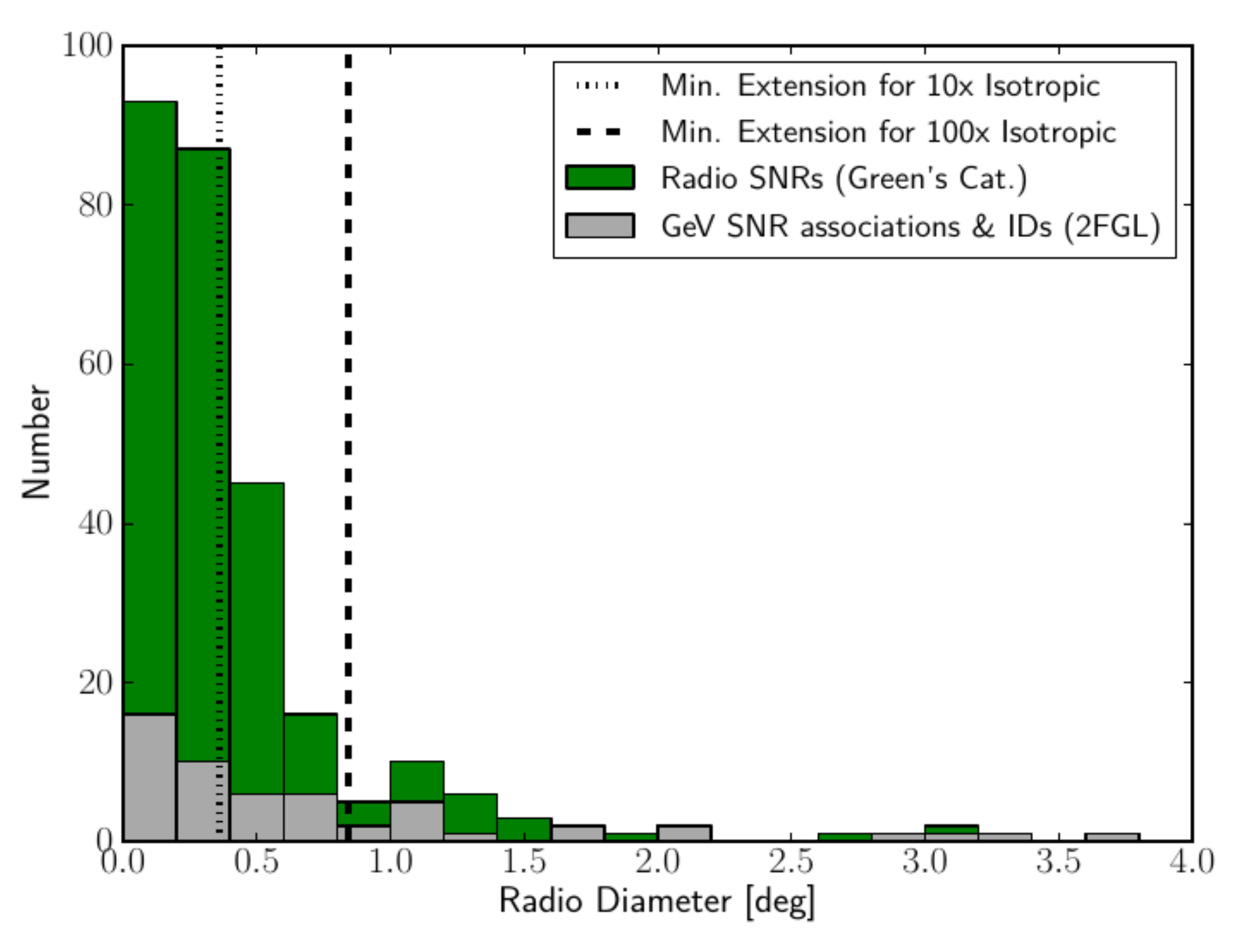} 
\end{overpic}
\caption{Distribution of SNR radio diameters from Green's catalog. The vertical dashed lines, the thin-dotted and thick-dashed lines indicate respectively the minimum detectable extension using $1\times$, $10\times$ and $100\times$ the isotropic background level, for more details see \cite{Lande12-extSrcSearch} and \cite{snrcat}. Roughly one third of the known Galactic SNRs may be resolved by the LAT if they are sufficiently bright GeV sources.}
\label{fig:size_hist_p}
\end{figure}

If GeV and radio sizes are similar, as has been observed on an individual basis for several extended SNRs (e.g. \cite{Lande12-extSrcSearch}), the LAT has sufficient spatial resolution to detect many SNRs as extended. Figure~\ref{fig:size_hist_p} shows the distribution of radio diameters from Green's catalog. Vertical dashed lines show the minimum detectable extension for source with flux and index typical of those observed in this catalog, based on simulations using the P7V6 instrument response functions (IRFs) \cite{Lande12-extSrcSearch}. 

\section{Detection method and source classification}\label{Sec:DetectMethod}
\begin{figure}[ht]
  \centering
  \begin{overpic}[width=0.56\columnwidth]{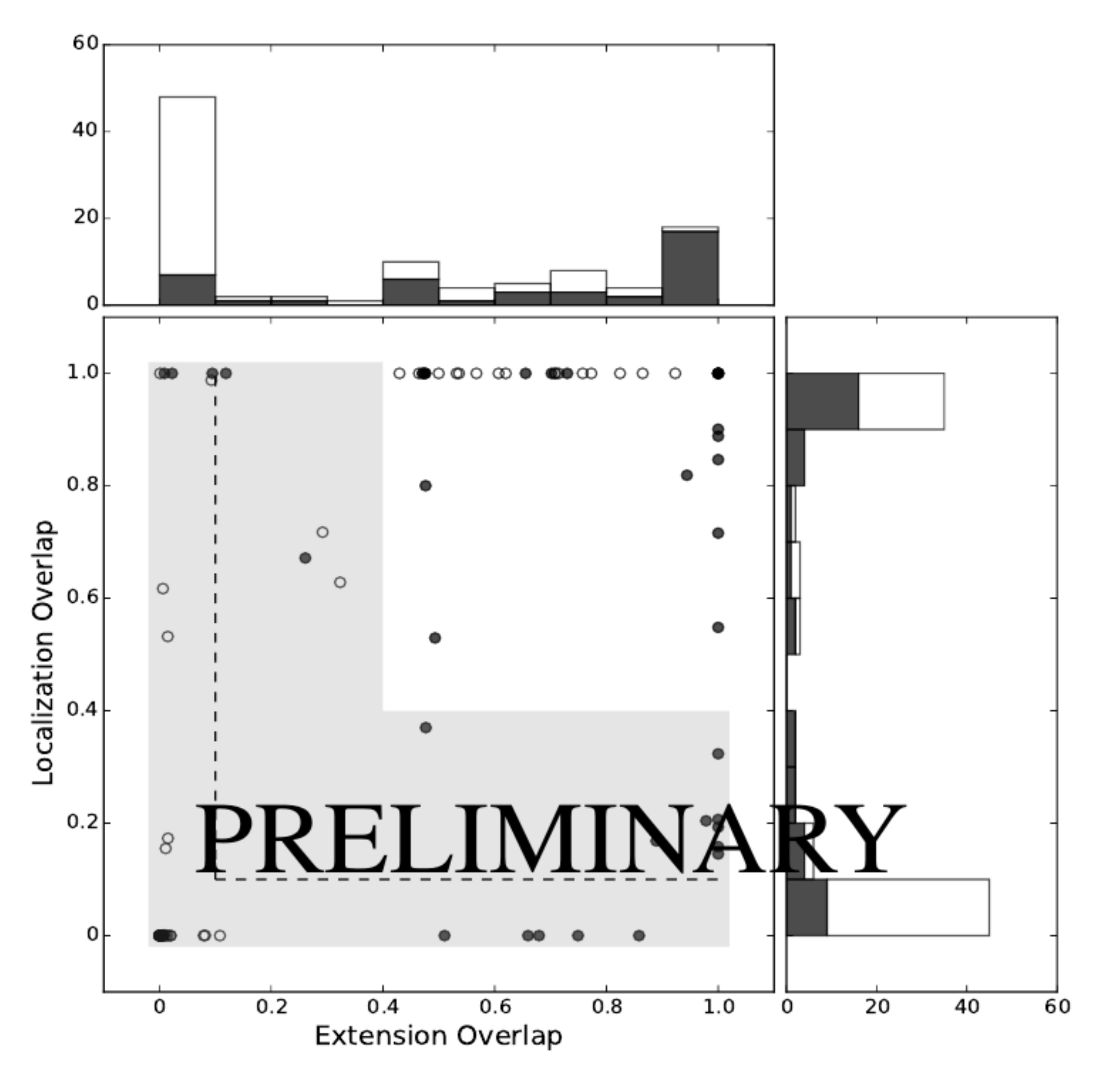} 
  \end{overpic}

  \caption{Distribution of $\overlaploc$ and $\overlapext$ for all significant GeV sources. GeV candidates with significant extension are shown as open circles and GeV candidates consistent with the point hypothesis are filled. Stacked bar histograms of both parameters are also shown. The classified candidate region discussed is shown with a white background. Those points in the grey region to the upper right of the dotted line are marginally classified candidates.}
  \label{fig:OverlapLocExt}
\end{figure}

For each SNR, we character size, morphology and spectrum of any \g-ray emission that may be coincident with the radio position reported in Green's catalog. This was achieved by testing multiple hypotheses for the spatial distribution of \g-ray emission: a point source and two different extended disk. One was obtained using only a disk source coincident in size with the radio size, while the second was obtained allowing the nearby background sources to be included in the final size of the SNR. The best fit was selected based on the global likelihoods of the fitted hypotheses and their numbers of degrees of freedom. 
To define association probabilities for the candidates, we compared radio positions and sizes available from Green's catalog, with that of the GeV candidates' localizations, localization errors, and extensions measured in this catalog.
Using this set of spatial information, we derived two parameters. The first, $\overlaploc$, provides a quantitative measure of whether the GeV localization is within the SNR's angular extent in radio. This parameter, ranging from $0$ to $1$, is calculated as:
\begin{equation}\label{eq:overlaploc}
 \overlaploc=\frac{\overline{Radio \cap
GeV_\mathrm{loc}}}{\min(\overline{Radio},\overline{GeV_\mathrm{loc}})}
\end{equation}
where $Radio$ represents the SNR's radio disk and $GeV_{\rm loc}$ is the GeV $95\%$ error circle. The notation $\overline{X}$ represents the area of $X$. 
The second parameter, $\overlapext$, quantifies whether the GeV candidate's localization and extension is consistent with the location and extension of the radio SNR:
\begin{equation}\label{eq:overlapext_ext}
 \overlapext=\frac{\overline{Radio \cap
GeV_\mathrm{ext}}}{\max(\overline{Radio},\overline{GeV_\mathrm{ext}})}.
\end{equation}
where $Radio$ is again the SNR's radio disk and $GeV_{\rm ext}$ is the best fit GeV disk if the GeV candidate is significantly extended. 
If the GeV detection is consistent with a point source, we determined if the corresponding SNR's radio size was consistent with it 
by redefining the $\overlapext$ parameter as: 
\begin{equation}\label{eq:overlapext_pt}
 \overlapext=\frac{\overline{Radio \cap GeV_\mathrm{min}}}{\overline{Radio}}.
\end{equation}
where $GeV_\mathrm{min}$ is the minimum resolvable radius, $GeV_\mathrm{min} \equiv 0.2^\circ$ for this analysis, see also \cite{Lande12-extSrcSearch}. 
The values of $\overlaploc$ and $\overlapext$ for all significant GeV sources are shown in Fig. \ref{fig:OverlapLocExt}.
We label the GeV detections with the most likely chance of true association as ``classified candidates,''
defined as those sources with $ \overlapext > 0.4$ and $\overlaploc > 0.4 $. 
``Marginally classified candidates'' are those GeV sources with a moderate chance of true association, 
defined as $\overlapext > 0.1$ and $\overlaploc > 0.1$ and at least one overlap estimator $<0.4$. Candidates which do not have overlap parameters in these categories are referred to as ``other'' sources. 
\section{Chance Coincidence}\label{Sec:ChanceCoinc}
In order to estimate the probability that any particular coincidence occurs by chance, we created a mock SNR catalog derived from Green's catalog with SNR positions randomized excluding the original location and we run our standard pipeline and classification method. Any classified source found would be by chance. 

After the analysis, we find that only \nmock~out of \nGalSNRs~mock SNRs are spatially coincident with a GeV excess. Comparing the $N_{\rm mock}=$\,\nmockN~mock coincidences to the $N_{\rm Green}=$\,\nassocprobclassified~candidates passing the association probability we estimate a false discovery rate of \nmockpercent{} for this specific realization of the mock catalog. 
We determined that, at $95\%$~confidence, the number of false discoveries will be less than~\nninetyfivemock{} for any mock catalog prepared as described above, corresponding to an upper limit of~\nninetyfivemockpercent{} for the false discovery rate. 

\section{Source of systematic errors}
For sources with significant emission, we estimated the systematic error propagating from the systematic uncertainty of the effective area and from the choice of IEM. For the former,
we estimated the systematic error by calculating bracketing IRFs, following the standard method \cite{LAT-instrument-paper}.
While for the choice of the IEM we developped a new method. An earlier version of this method was described in \cite{dePalma13-AltIEMSystematics_FSymp}.
Interstellar emission contributes substantially to LAT observations in the Galactic plane, where a majority of SNRs are located. Moreover, interstellar \g-ray emission is highly structured on scales smaller than the regions of interest (RoIs) typically used for this analysis. To explore the systematic effects on SNRs' fitted properties caused by interstellar emission modeling, we have developed a method employing eight alternative IEMs. 
The work in \cite{Ackermann12-aIEMs}, using the GALPROP\footnote{\url{http://galprop.stanford.edu/}} cosmic ray (CR) propagation and interactions code, was the starting point for our alternative IEM building strategy. This strategy is different from what was adopted in the development of the standard IEM used in the rest of the analysis and in the usual \FermiLat~analysis. 
In order to build those eight models, we varied the values of three input parameters that were found to be the most relevant in modeling the Galactic plane: CR source distribution, height of the CR propagation halo, and \hi{} spin temperature \cite{Ackermann12-aIEMs}. 
The models were constructed to have separate templates for emission associated with gas traced by \hi{} and CO in four Galactocentric rings and an inverse Compton (IC) template covering the full sky. By allowing separate scaling factors for these different components of the model, we allowed many more degrees of freedom in fitting the diffuse emission to each RoI.

\begin{figure}[ht]
  \centering
  \begin{subfigure}{.45\textwidth}
  \centering
  \begin{overpic}[width=\linewidth]{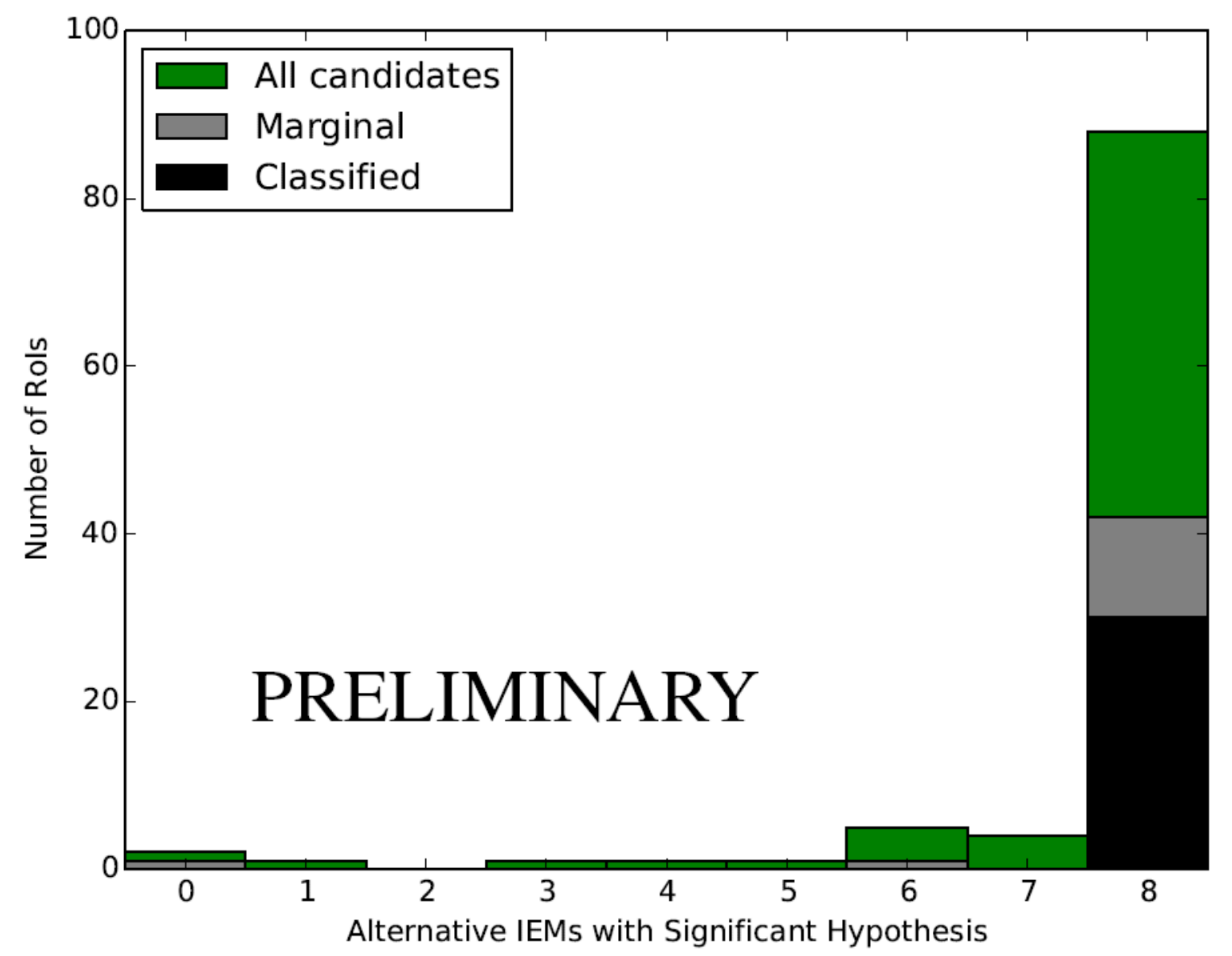}
  
  \end{overpic}
  \subcaption{Significance comparison\label{fig:besthyp_sig}}
  \end{subfigure}
  \begin{subfigure}{.44\textwidth}
  \begin{overpic}[width=\linewidth]{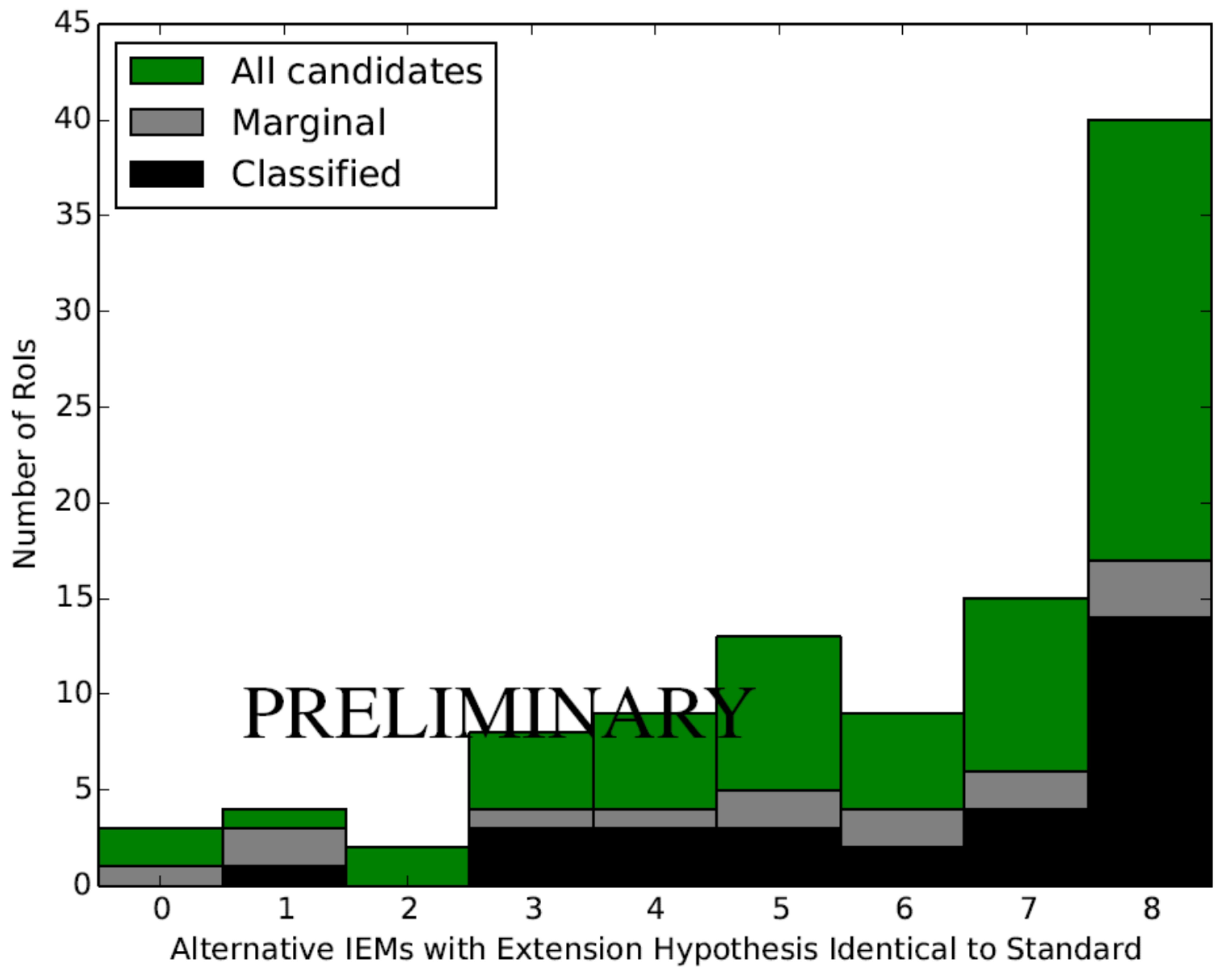}  
  \end{overpic}
  
  \subcaption{Best-fit model comparison \label{fig:besthyp_sig_ext}}
\end{subfigure}
   \caption{For each candidate's analysis with each of the alternative IEMs, we computed the candidate's significance and the best extension hypothesis, either point-like or extended. {\it Left:} Number of analyses of each region with each of the alternative IEMs for which the candidate remained significant (Test Statistic$>9$). {\it Right:} Number of analyses of each region for which the best extension hypothesis with the alternative IEM remained the same as that found with the standard one.}
\end{figure}

For each significant candidate SNR we considered two hypotheses: the point source and one of the extended hypothesis that is preferred in the main pipeline analysis. For each significant candidate SNR we performed independent fits for each hypothesis for each of the eight alternative IEMs as well as the standard IEM, for a total of $18$ fits of the each RoI. For each candidate SNR and alternative IEM we evaluate the preferred extension hypothesis, shown in figure \ref{fig:besthyp_sig}, and its significance, in figure \ref{fig:besthyp_sig_ext}.

\begin{figure}[ht]
  \centering
  \begin{subfigure}{.45\textwidth}
  \centering
  \begin{overpic}[width=\linewidth]{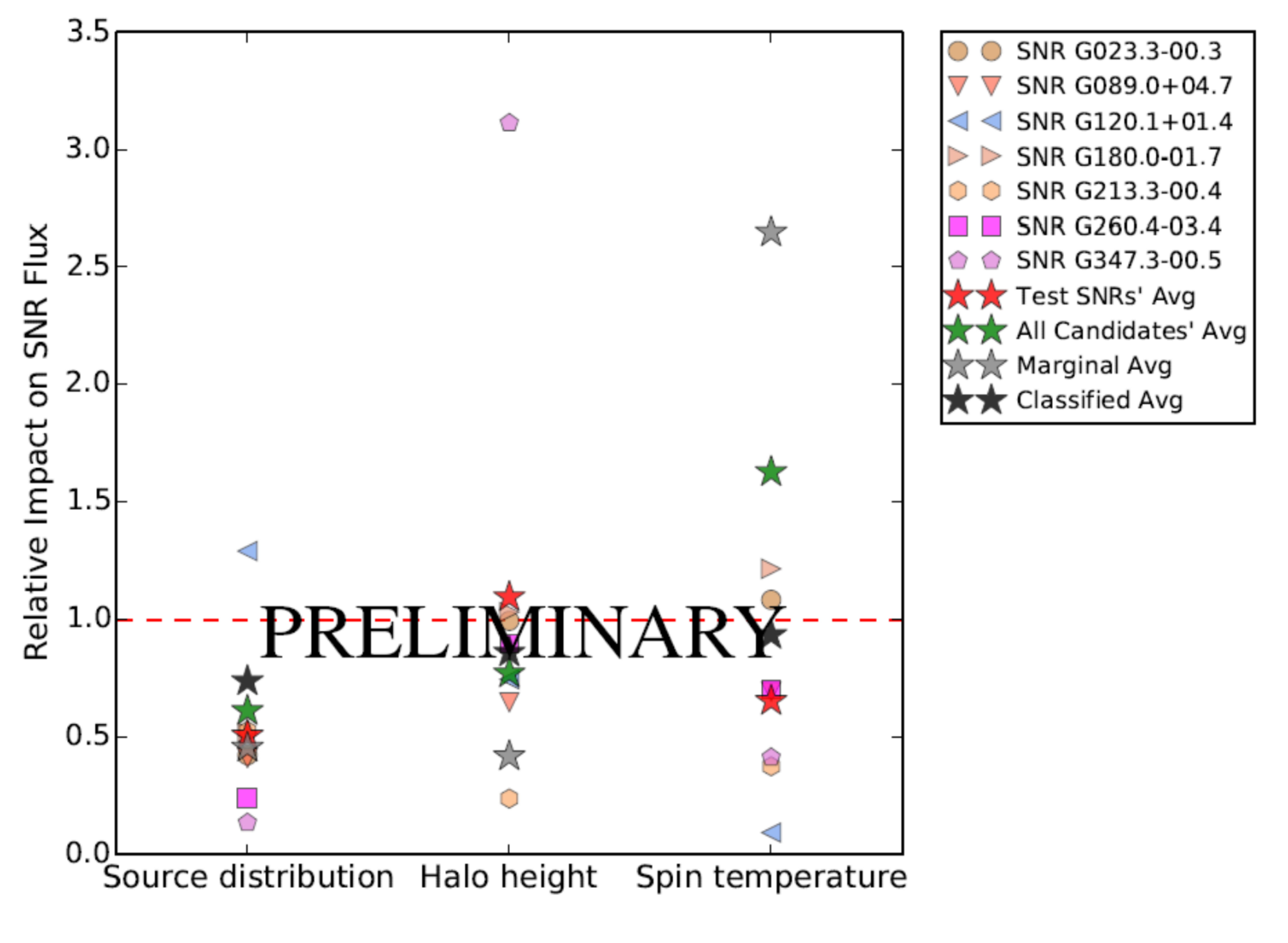}
  
  \end{overpic}
  \subcaption{Flux \label{fig:input_param_aIEM_flux}}
  \end{subfigure}
  \begin{subfigure}{.45\textwidth}
  \begin{overpic}[width=\linewidth]{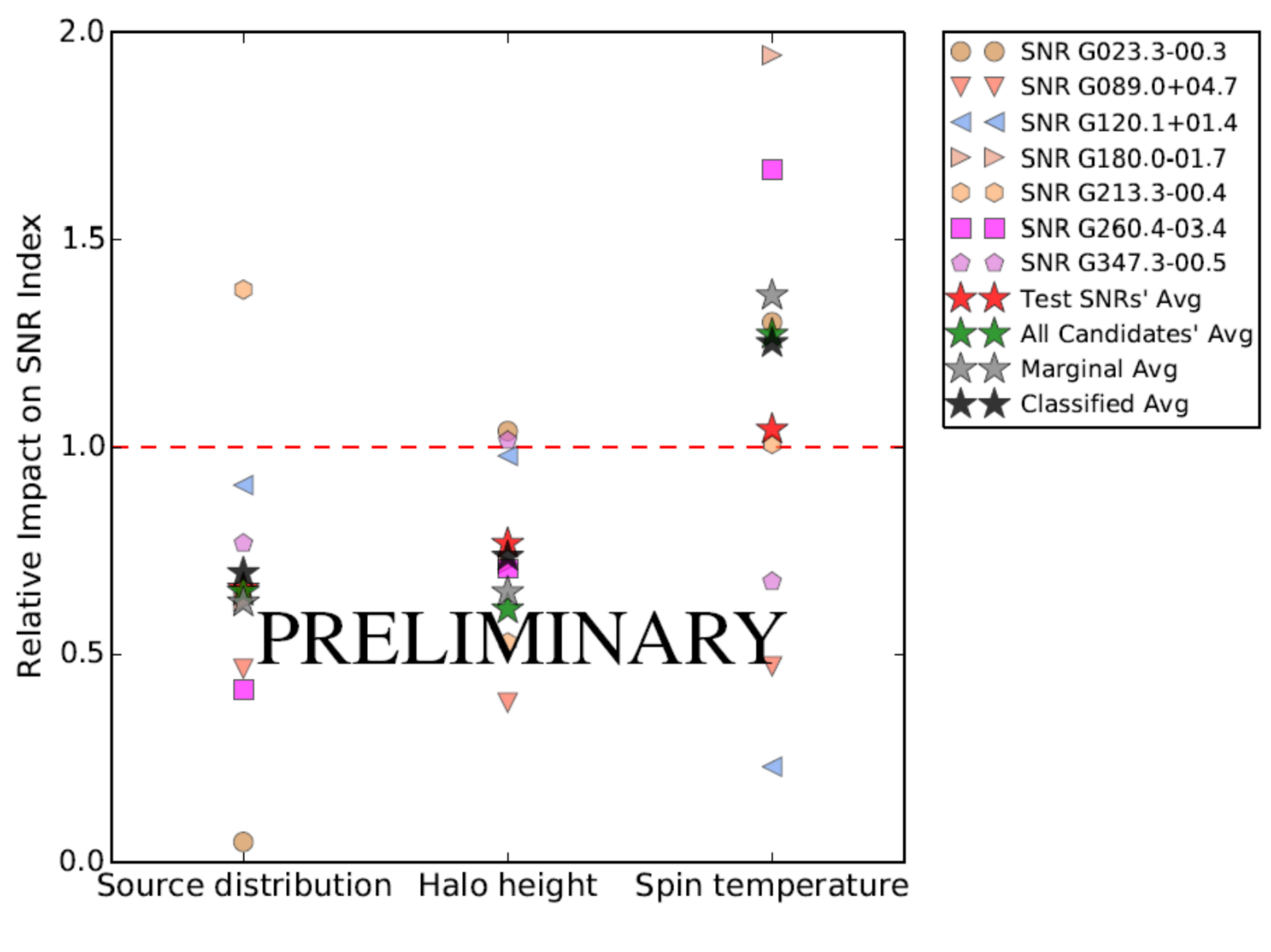}  
  \end{overpic}
  \subcaption{Index \label{fig:input_param_aIEM_Index}}
\end{subfigure}
   \caption{The impact on the fluxes and on the PL index of candidates for each of the alternative IEM input parameters (source distribution, halo height, and spin temperature), marginalized over the other GALPROP input parameters, is shown ($R$ in the text). The stars represent the average ratio over the different candidates' sets (classified, marginal, test and all). As no alternative IEM input parameter has a ratio significantly larger than $1$, no input parameter dominates the systematic uncertainties of the fitted source parameter sufficiently to justify neglecting the others. }
\end{figure}

To identify which, if any, of these three sets of GALPROP parameters (CR source distribution, height of the CR propagation halo, and \hi{} spin temperature) has the largest impact on the fitted source parameters, for all the RoIs we marginalized over the other two parameter sets and examined the ratio of the averaged candidates' parameter values relative to the values' dispersion. For a fitted source parameter~$P$, such as flux, and a GALPROP input parameter set $P_\mathrm{aIEM} = \{i, j\}$, e.g. spin temperature $T_s~=~\{150$\,K$, 10^5$\,K$\}$, the ratio is:
\begin{equation}
\label{eq:ratio_aIEM}
R \equiv \frac{|<P_{i}>-<P_{j}>|}{max(\sigma_{P_i},\sigma_{P_j})}.
\end{equation}
In the previous definition $<P_{i}>$ and $\sigma_{P_i}$ are, respectively, the average and the standard deviation of the sample composed by the candidate parameters $P$ obtained after the fit with GALPROP input models with the parameter $i$ (e.g. spin temperature $T_s~=~150$). 
A value of $R >> 1$ for a particolar GALPROP input parameter set means that all the other sets are negligible for that particolar candidate. If in various RoIs different GALPROP input parameter sets (CR source distribution, CR propagation halo height and \hi{} spin temperature) have a larger effect on $P$ than the others, we need to test each source with all the alternative IEMs since in different part of the sky different maps are more relevant. As you can see in  Figure~\ref{fig:input_param_aIEM_flux} for the flux and \ref{fig:input_param_aIEM_Index} for the power-law (PL) index, none of the three GALPROP input parameters has $R$ significantly smaller or larger than $1$ such that the other GALPROP parameters can be neglected for all the candidates tested.

 We evaluated the systematic error due to the uncertainties of the interstellar medium modelling using the following formula. For each fitted parameter $P$ we obtain a set of $M$\,$=8$ values $P_i$ that we compare to the value obtained with the standard model $P_\mathrm{STD}$. Our estimate of the systematic uncertainty on $P$ due to the modeling of interstellar emission is:
\begin{equation}
\label{eq:sys_err_weighted}
 \sigma_\mathrm{sys,w}=\sqrt{\frac{1}{\sum^M_i w_i} \sum^M_i w_i(P_i-P_\mathrm{STD})^2}.
\end{equation}
In Equation ~\ref{eq:sys_err_weighted} the weights are $w_i=1/\sigma_i^2$ where $\sigma_i$ is the statistical error of a parameter with a particular alternative IEM.

 We note that this strategy for estimating systematic uncertainty from interstellar emission modeling does not represent the complete range of systematics involved. In particular, we have tested only one alternative method for building the IEM and varied only three of the input parameters. This ensemble of models therefore cannot be expected to encompass the full uncertainty associated with the IEM.
Further, as the alternative method differs from that used to create the standard IEM, 
the parameters estimated with the alternative IEM may not bracket the value determined using the standard IEM. Our estimate of the systematic error in Equation~\ref{eq:sys_err_weighted}, accounts for this.
 
\section{Error comparisons}
The systematic error on flux estimated from the alternative IEMs tends to dominate over that due to effective area. This is true for most sources, regardless of their size, classification, or Galactic longitude. Moreover, the error propagated from the effective area tends to be $<10\%$ of the flux for the large majority of candidates, while that from the alternative IEMs ranges over $\sim3$ orders of magnitude, indicative of the diverse environments in which these candidates are found. We also observe that the candidates classified as other than SNRs, particularly when extended, tend to have somewhat larger alternative IEM errors than the classified and marginally classified candidates.
For the PL spectral index the systematic errors from the bracketing IRFs dominate the alternative IEM errors for approximately half the candidates. Aside from candidates classified as other having somewhat larger systematic errors due to the alternative IEMs, systematic errors on neither flux nor index are significantly different among various subtypes: point or extended candidates, marginal or classified candidates, or young compared with interacting.

\begin{figure}[ht]
  \centering
  \begin{subfigure}{.49\textwidth}
  \centering
  \begin{overpic}[width=\linewidth]{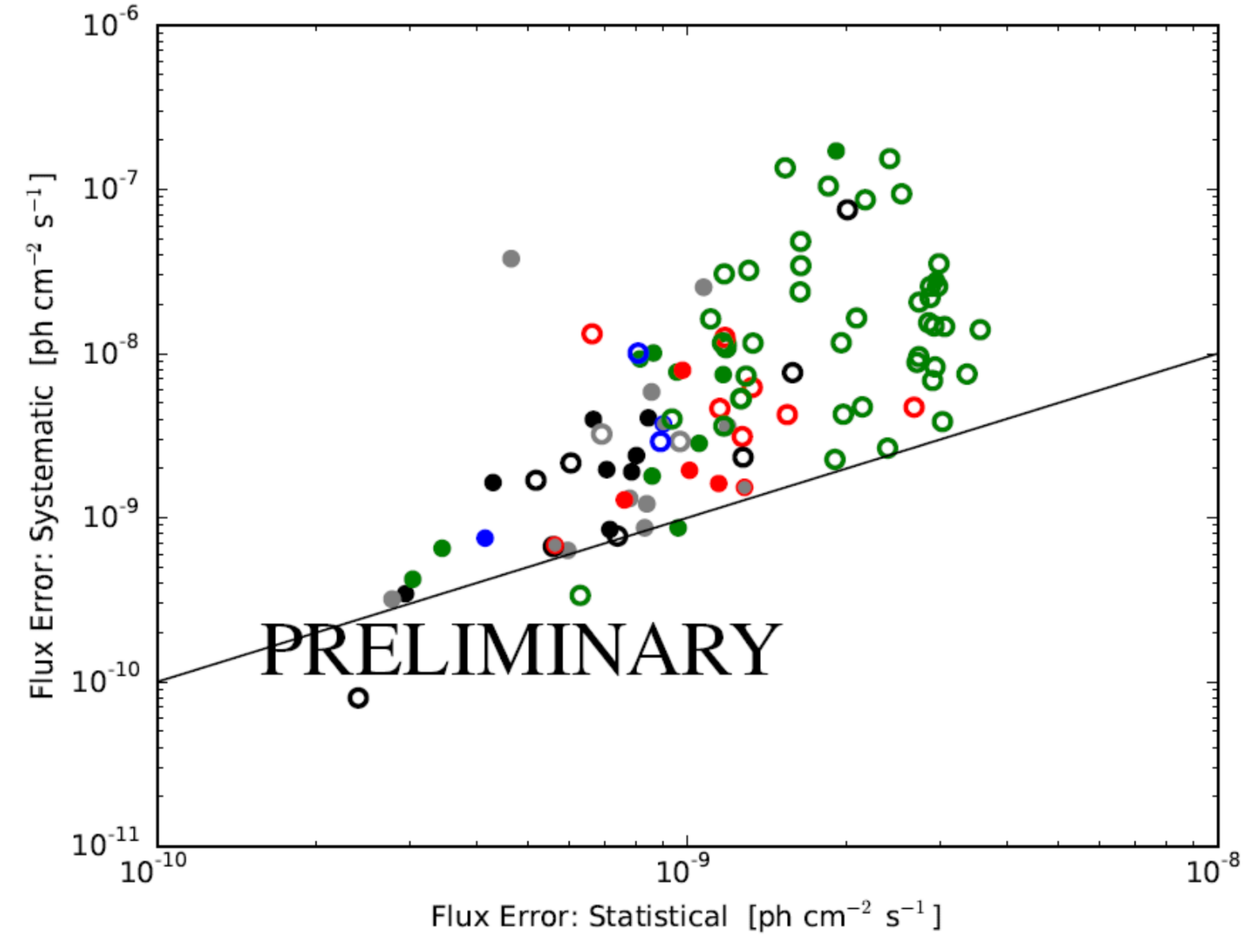}
  
  \end{overpic}
  \subcaption{Flux \label{fig:sysvStatFluxErrs}}
  \end{subfigure}
  \begin{subfigure}{.48\textwidth}
  \begin{overpic}[width=\linewidth]{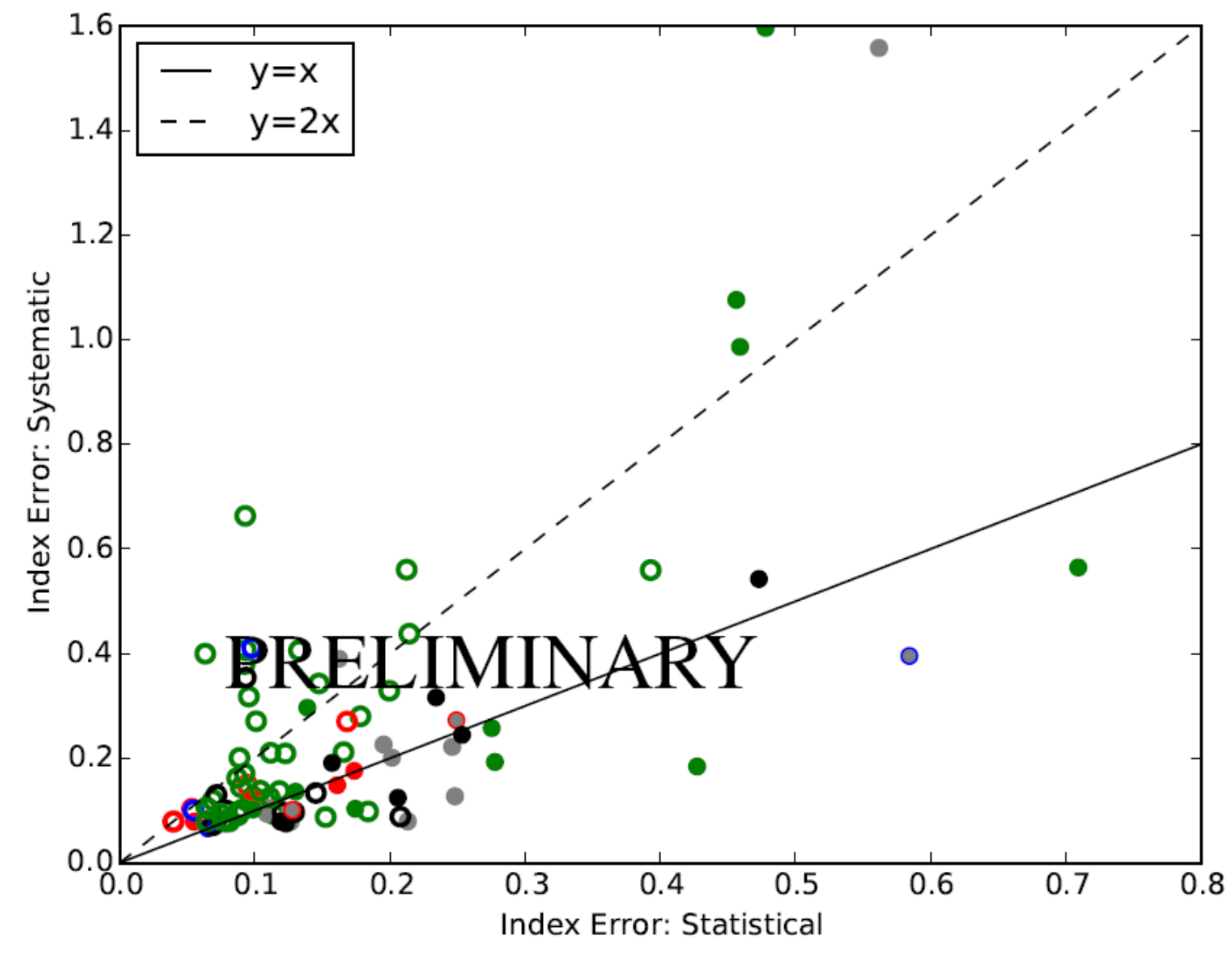}  
  \end{overpic}
  \subcaption{PL Index. \label{fig:sysvStatIndexErrs}}
\end{subfigure}
   \caption{Comparison of the statistical and systematic errors, the latter derived from the alternative IEMs and the bracketing IRFs, for the flux and PL index. Open circles indicate extended SNRs while filled circles indicate point-like sources. All SNRs that passed classification are shown as black unless also classified as young non-thermal X-ray SNRs (blue) or as interacting with MCs (red). Candidates which did not pass classification but still had both fractional overlaps $>0.1$ are grey. If they are also young or interacting, they are outlined in blue or red, respectively. with the addition that all candidates classified as ``other'' are shown in green. For the flux the systematic error typically dominates the statistical error on the flux for all classes of candidates. For the index, in a number of cases the statistical error dominates the systematic.}
\end{figure}

The total systematic error, derived adding in quadrature the bracketing IRFs and alternative IEMs errors, dominates the statistical error on a candidate's flux, as seen in Figure~\ref{fig:sysvStatFluxErrs}. The statistical errors for PL index dominate over systematics for a number of candidates as shown in Figure~\ref{fig:sysvStatIndexErrs}.

\section{Conclusions}
We detected \ndetected~candidates with a final source TS $>25$ in the \nGalSNRs~SNR RoIs (see Section~\ref{Sec:DetectMethod}). Of the \ndetected~detected candidates, \nassocprobclassified{} passed the association probability threshold. Of these, \nclassifiedsnrs~SNRs ($\sim11\%$ of the total) show significant emission for all alternative IEMs and are classified as likely GeV SNRs. An additional \nnotsnrs{} were identified as sources which are not SNRs; \nmarginalAIEM~other candidates were demoted to marginal due to their dependence on the IEM.
Of the sources likely to be GeV SNRs, \nextended~show evidence for extension (TS$_{ext} > 16$). Only sources associated with SNRs G34.7$-$0.4 and G189.1+3.0 show evidence of significant spectral curvature in the $1-100$\,GeV range and are fit with logP spectra. 
Of the classified candidates, \nnewextended~extended and \nnewpointlike~point SNRs are new. For all the sources we evaluated their spectral and spatial characteristics with systematic and statistical uncertainties. A description of the catalog results in a multiwavelenght contest and their implication in the understanding of the CR acceleration can be found in \cite{snrcat} and \cite{ICRC_talk}.

\section{Acknowledgments}

 The \textit{Fermi}-LAT Collaboration acknowledges support for LAT development, operation and data analysis from NASA and DOE (United States), CEA/Irfu and IN2P3/CNRS (France), ASI and INFN (Italy), MEXT, KEK, and JAXA (Japan), and the K.A.~Wallenberg Foundation, the Swedish Research Council and the National Space Board (Sweden). Science analysis support in the operations phase from INAF (Italy) and CNES (France) is also gratefully acknowledged.

\end{document}